\documentstyle[12pt,epsf,epsfig]{article}

\def\be{\begin{eqnarray}}
\def\ee{\end{eqnarray}}

\def\pd{\partial}
\newcommand{\f}[2]{\frac{#1}{#2}}
\newcommand{\partl}[2]{\frac{\partial #1}{\partial #2}}

\begin{document}

\title{Renormalization Group Flow Equation at Finite Density}

\author{%
J. Meyer$^{1}$,
G. Papp$^{1,2,}$\thanks{On leave from HAS Research Group for Theoretical
Physics, E\"otv\"os University,  P\'azm\'any P. s. 1/A, Budapest, H-1117
Hungary},
H.-J. Pirner$^{1,3}$
and
T. Kunihiro$^{4}$ 
 \\[2mm]
{\small $^1$ Institut f\"ur Theoretische Physik der
Universit\"at Heidelberg,}\\ {\small D-69120 Heidelberg, Germany} \\
{\small $^2$ CNR,
Department of Physics, Kent State University, 44242 OH USA} \\
{\small $^3$ Max-Planck-Institut f\"ur Kernphysik, D-69029 Heidelberg, 
        Germany} \\
{\small $^4$ Faculty of Science and Technology,  Ryukoku University, Seta,
  Ohtsu 520-2194, Japan}
}

\date{\today}

\maketitle
\begin{abstract}
For the linear sigma model with quarks we derive  
renormalization group flow equations for finite temperature and
finite baryon density using the heat kernel cutoff. At
zero temperature we evolve the effective potential to the Fermi momentum and compare
the solutions of the full evolution equation with 
those in the mean field approximation. We find
a first order phase transition either from a massive constituent quark phase
 to a mixed phase, where both massive and massless quarks are present,
or from  a metastable constituent quark phase at low density to a 
stable massless quark phase at high density.
In the latter solution, the formation of droplets of massless quarks
is realized even at low density. 
\end{abstract}


\newpage
\section{Introduction}
Recently a  transparent approach to the evolution of a hadronic system
with resolution has been constructed using a Schwinger proper time
representation
of the fluctuation determinants of quarks and chiral mesons~\cite{BJS}.
This method gives a  picture of the transformation of constituent
quarks at low resolution into partonic massless quarks at high resolution
observed in deep inelastic scattering
~\cite{Pirner}.
At finite temperature the same type of renormalization group (RG) flow
equations  
give critical indices for the chiral phase transition in agreement with 
the $O(4)$ model~\cite{BJS,BJS2}. Present finite temperature QCD lattice
simulations seem to indicate such an $O(4)$ type behavior, with some
uncertainty.

Simulation of QCD at finite baryon density on a lattice 
is still a challenge:
 In Euclidean space the 
chemical potential gives rise to a complex action which forbids
Monte Carlo calculations. Quenched simulations at finite density suffer
from additional  shortcomings~\cite{MISA}. The challenge  is
therefore to come up with a
calculational scheme for finite baryon density which has a well controlled
predictive power.
Theoretical studies of high density matter given in this work
 are indispensable to
 understand what is going on with high-energy
 heavy ion collisions which  probe not only high excitation
energy (perhaps temperature) but also high baryonic density. 
Recent studies have suggested a very rich phase structure at high baryon
density~\cite{DIQUARK1,DIQUARK2}. There has been considerable work
devoted to extract the equation of state of nuclear matter in terms
of nucleon - nucleon potentials.  The problem is to link
the high density region accessed by high energy heavy ion collisions with
the low density nuclear physics region.

At low density it is probably not very efficient to describe
baryonic matter by quarks, but in an
intermediate region, quarks and mesonic bound states may be the 
right degrees of freedom. The mesons are 
formed by 
the strong gluonic attraction in certain channels. The
chiral linear sigma model with quarks is a good model 
to investigate
the dynamics below a momentum scale of $k<k_{uv}=1.2 \;
{\rm GeV}$\footnote{The definition of this momentum scale depends on
the renormalization group scheme.}. Such a hybrid approach has been
already used at finite temperature with success.  
%
%
In this paper we will treat pure
quark matter and concentrate on chiral
symmetry aspects of the transition. We neglect quark confinement but assume
the gluons to be confined into the mesons.
The connection to nuclear matter will be considered
elsewhere~\cite{Schwenzer}. The finite
baryon density dynamics is
much more sensitive to neglecting confinement
than the finite temperature dynamics.  Indeed it is
known that e.g. the \emph{NJL} model gives a first order phase transition
at unrealistically low baryon density~\cite{Huef,Buballa}.
Without confining quark forces, which
repel quarks tripled in neutral nucleons from each other~\cite{confin},
the linear
sigma model with quarks probably overestimates the binding energy per
baryon of nuclear matter.

We use renormalization group flow equations in the presence of finite
quark density. As a first step, we start with the heat kernel
representation of the effective potential using the cut-off function
of the earlier finite temperature calculation~\cite{BJS}. 
The extension of this technique to finite density is straightforward
in contrary to the cut-off function introduced by the Wetterich
group~\cite{Wett} for finite temperature studies. However, the proposed
method has its shortcomings: {\bf 1.} The
heat kernel cutoff is a Lorentz invariant cutoff, and as such does not
zoom to the Fermi surface with increasing evolution steps. In principle
the goal of the renormalization group approach for finite fermion
density is~\cite{Pol} to treat the critical long wave length particle
hole excitations of the finite Fermi system towards the end of the
evolution. These particle hole excitations are generated at the Fermi
surface.  In the relativistic case with Goldstone bosons this final
evolution step should coincide with the treatment of the zero mass
bosons.
{\bf 2.} Extra terms which reduce the ultraviolet sensitivity in our
heat kernel cut-off function result in an integration pole at the Fermi
surface, making the numerical integration difficult.

In this paper we concentrate on deriving the formalism for a
finite density calculation
with finite temperature. We solve the flow equations for
finite density at temperature $T=0$ leaving our 
final goal, the determination of the phase diagram in the whole $(T,\mu)$
plane to another paper. Furthermore, we compare our renormalization
group results with the mean field approximation.

As a description of choice in this work we use the linear sigma
model. We think that this model can play the role of a 
simplified theory showing the effects of
relativistic field theory at finite density as a kind of 
Ising model for nuclear physics. 
The equation of state improved by the renormalization group in the one loop 
gives clear  signals on what
is happening in the finite  density system. The main message is the
intricate connection between the effective mass of the fermions and the
effective masses of the mesons exchanged between the fermions. 
In classical nonrelativistic nuclear physics the time scale of the
meson exchange is short in comparison with the time scale of the nucleon
dynamics. At normal nuclear density the mesons will not be modified
drastically by the nuclear medium.
In the quark  picture the time 
scale for the exchange of a meson is not different from the time scale
connected to the rearrangement of  the fermion momenta. The mesons
cannot be taken into account as potentials, which are turned into effective
interactions, due to the quark rescattering in the medium. The mesons
participate fully in the reorganization of the chiral symmetric phase with
increasing density. The quark many body problem is therefore very
different from the nuclear physics problem where
the nucleon dynamics largely decouples form the meson dynamics.
The only possible tool we have to solve such an intrinsically nonlinear
problem in field theory is the renormalization group, therefore
it is worth to study this technique in nuclear physics to get a better
understanding of the phase structure of baryonic matter.

The paper is organized as follows: 
in section 2 we review the derivation of the renormalization group
flow equation for the effective potential at zero density and
temperature. The extension of this method to finite temperature and
baryon density is presented in section 3. At the end of this section we show a
first result from the RG flow equation at finite density. In section 4 we discuss the
mean field approximation to the linear sigma model and in section 5 we
present our results for two sets of mean field couplings, obtained from
the RG evolution. Here we study the finite baryon density phase
transition of the linear sigma model in the mean field approximation.
In section 6 we compare the results of the grid and the mean field
calculations.
Section 7 is devoted to a summary and to the conclusions.
In the Appendices we discuss the connection between the fermionic part of the
flow-equations and the mean field result and give a detailed derivation
of the flow equations at finite temperature and density. 

\section{Evolution of the linear $\sigma$ model}
\label{sec:evol-lin-sig}
We consider the chiral constituent quark model with quarks, $\sigma$ and
$\pi$ mesons.
At zero temperature and chemical potential in Euclidean space the
partition function or the generating
functional without external sources $\Delta$, is given by
\be\label{zz1}
Z[\Delta=0] & = & \int {\cal D}q {\cal D}\bar{q} 
{\cal D}\sigma  {\cal D}\vec{\pi}
 \exp\{-\int d^4x \left( {\cal L}_F + {\cal L}_B \right) \}
\ee
with a fermionic ${\cal L}_F$, and a bosonic ${\cal L}_B$, parts,
\be\label{fermil}
{\cal L}_F & = & \bar{q}(x) \left[ \gamma_E \partial_E +
g\left( \sigma + i\vec{\tau}\vec{\pi}\gamma_5 \right)\right] q(x)\,,
\ee
\be\label{bosel}
{\cal L}_B & = & \frac{1}{2} \left[
(\partial_\mu \sigma)^2 + (\partial_\mu \vec{\pi})^2 \right] +
F(\sigma^2\!+\!\vec{\pi}^2)\,.
\ee
The Yukawa coupling of the constituent quarks to the mesons is
denoted by $g$. 
The parameters of the linear $\sigma$-model
at $T=0$ are chosen in the same way as in ref. \cite{BJS,BJS2}.
We assume that at an ultraviolet scale $k_{uv}$, 
the full $QCD$ dynamics reduces
to a hybrid description in terms  of quarks and chiral bound states. 
%
%
The gluons are assumed to be frozen in the residual mesonic degrees of
freedom and their couplings. 

At the beginning of the evolution at 
$k_{uv}=1.2 \; {\rm GeV}$ we choose
the effective potential density $F_0$, to be of the following form:
\begin{eqnarray}
\label{eq:pot}
F_{0}(\vec \phi) &=& \frac{m^2}{2}\phi^2 +\frac{\lambda}{4}\phi^4 \,,
\qquad\mbox{with}\quad \phi^2 
        = \sigma^2\!+\!\vec \pi^2,
\end{eqnarray}
where the positive mass squared $m^2=0.4^2 \; {\rm GeV}^2$ reflects a symmetric
ground state, i.e.  the minimum of the potential lies at the origin.  The four
boson coupling at this scale is $\lambda=30$~\cite{BJS,BJS2}. The values correspond to a
critical temperature $T_c\approx 150$ {\rm MeV}, and a chiral symmetry breaking
scale $k_{br}\approx 1$ {\rm GeV} which one obtains after performing the RG
procedure.   

The effective potential density $F(\phi)$, can be evolved ~\cite{BJS} using
the heat kernel method. For this
purpose the one loop effective potential is calculated with a cutoff
function $f(k^2\tau)$, which contains the evolution scale $k$:
\be
\label{eq:cutoff}
f(x=k^2\tau)= e^{-x}\ \left(1\!+\!x\!+\!\f12 x^2\right).
\ee
Doing this, the couplings of the of the effective potential become scale
dependent: $m = m(k)$ and $ \lambda = \lambda(k)$.

Using the the Schwinger proper time representation the fermionic $F^F$
contribution to the effective potential density is:
\be
  \label{eq:uf}
  F^F&=& \f{1}2 \int\limits_0^\infty\! \f{d\tau}{\tau}\ 
        f(k^2\tau) \int \!\frac{d^4q}{(2 \pi)^4} \mbox{Tr}\ 
          e^{-\tau[q^2+g^2\phi^2]} \\
  &=& \f{1}2 4N_c N_f \int\limits_0^\infty\! \f{d\tau}{\tau}\  
        f(k^2\tau) \int \!\frac{d^4q}{(2 \pi)^4} 
                e^{-\tau[q^2+g^2\phi^2]} \nonumber
\ee
and correspondingly~\cite{BJS} the bosonic part is: 
\be
  \label{eq:ub}
  F^B&=& -\f{1}2 \int\limits_0^\infty\! \f{d\tau}{\tau}\ 
        f(k^2\tau) \int \! \frac{d^4q}{(2 \pi)^4}  \mbox{Tr}\ 
        e^{-\tau[q^2+\f{\partial^2 F_0}{\partial \phi_i
                \partial \phi_j}]} \\
  &=& -\f{1}2 \int\limits_0^\infty\! \f{d\tau}{\tau}\ 
        f(k^2\tau) \int \! \frac{d^4q}{(2 \pi)^4}  e^{-\tau[q^2]} \left[
                3 e^{-\tau 2 F_0^\prime} + 
                e^{-\tau[2 F_0^\prime +4 F_0^{\prime\prime} \phi^2]} \right],
        \nonumber
\ee
with
\begin{equation}
  \label{eq:deriv_F}
  F_0 = F_0(\phi^2,k), \;\;\;
  F_0^\prime = \frac{\partial F_0}{\partial \phi^2}, \;\;\;
  F_0^{\prime\prime} = \frac{\partial^2 F_0}{(\partial \phi^2)^2}.
\end{equation}

The total effective potential
density is the sum of these two terms,  $F=F^B+F^F$.
The evolution equation of this potential results from the derivative of the
potential with respect to $k$. In the spirit of the
renormalization group improved 
1-loop approximation the derivative only acts on the cutoff function
and the potential density $F_0$, is replaced by the evolving potential
density $F$. 
The evolution equation for the linear sigma model then has the following
simple form:
\be
  \partl{F}{k} &=& 
        \displaystyle \f{k^5}{32\pi^2} \left\{
        \f3{k^2\!+\!2F^{\prime}}\!+\!
        \f1{k^2\!+\!2F^{\prime}\!+\!4 F^{\prime\prime}\phi^2} 
        \!-\f{8N_c}{k^2\!+\!g^2\phi^2}  \right\}.
\label{eveq}
\ee\
In the approximation to the RG-evolution used here only the effective
potential density $F$, evolves with the scale $k$. The Yukawa-coupling
$g=3.23$ is assumed to be constant.  The limitation of this approach
will be discussed later.  
%
%
Note that during the evolution one passes from
the region with $m^2 > 0$, where the potential is symmetric, to the region $m^2 <
0$ where the potential has a mexican hat shape. 
The denominators of the meson loop terms (e.g. the first two terms on the
l.h.s. of eq.~(\ref{eveq})) indicate a limitation
of the one loop renormalization flow. The one
loop corrections are of order ${\cal O}(\hbar)$ in an
expansion in $\hbar$. In the regime
$k^2\!+\!2F^{\prime}\leq 0$ however, the usual Gaussian fixed point is
unstable and a non-Gaussian fixed point should be considered bringing in
${\cal O}(\hbar^0)$ effects~\cite{Janos}. With our choice of parameters
this happens at $k\leq k_{inst}\approx 200$ {\rm MeV}.


\section{Evolution equation at finite density}

The renormalization group flow equations give a well determined 
shape of the effective potential with which we will work in this
section. Since they are formulated in the continuum, the finite baryon density
$\rho_B$ can be implemented. Although the primary aim of the present work is
to discuss the chiral transition at finite baryonic density, we shall present
the master formulae for finite temperature $T$ and finite $\rho_B$ and then
take the limit $T \rightarrow 0$. 

We start our derivation from eqs.~(\ref{eq:uf},\ref{eq:ub}) extending
them to finite temperature and 
finite chemical potential. Since we are working now with fixed
temperature and chemical potential, one should replace the potential
density $F$, by the thermodynamical potential density $\Omega(\phi^2)$.
The fermionic part of the effective potential density generalized to
finite temperature $T$ and  chemical potential $\mu$ is
\be
\label{omferm}
\Omega^F = \f{T}2 4N_c N_f \int\limits_0^\infty\! \f{d\tau}{\tau}\  
        f(k^2\tau) \sum_{n} \! \int \!\! \frac{d^3 \! q}{(2\pi)^3} 
                e^{-\tau[(\nu_n+i\mu)^2+\vec{q}^2+g^2\phi^2]},
\ee
with the cutoff function $f(x)$ given by eq.~(\ref{eq:cutoff}), 
and the Matsubara frequencies $\nu_n=(2n+1)\pi T$ for fermions. 
The bosonic part is the one already examined in~\cite{BJS}, 
\be
\Omega^B = -\f{T}2 \int\limits_0^\infty\! \f{d\tau}{\tau} \! 
        f(k^2\tau) \sum_{n} \! \int \!\!\! \frac{d^3 \! q}{(2\pi)^3}
                e^{-\tau(\omega_n^2+\vec{q}^2)} \!\left[
                3 e^{-\tau 2 \Omega_0^\prime} + 
                e^{-\tau(2 \Omega_0^\prime 
+4 \Omega_0^{\prime\prime} \phi^2)} \right]
\ee
with Matsubara frequencies $\omega_n=2n\pi T$ for bosons.
The derivatives in the potential are taken again 
with respect to $\phi^2$. The total effective potential
density is the sum of the two terms,  $\Omega=\Omega^B+\Omega^F$.

An advantage of the heat kernel regulator is that 
the calculation of derivatives with respect to the momentum scale $k$
can be performed analytically to yield  compact formulae.
As shown in appendix A, we have 
\be
\label{eq:fermi-1}
\frac{\pd \Omega^F}{\pd k^2}=\frac{N_cN_f}{8\pi^2}k^4
       \frac{d}{dk^2} \int_0^{\infty}dq
    \frac{1}{E_{q,k}}\left[1-n(E_{q,k})-\bar{n}(E_{q,k})\right],
\ee
where $E_{q,k}=\sqrt{q^2+k^2+g^2\phi^2}$ and
 $n(x)$ and $\bar{n}(x)$ are the Fermi-Dirac distribution functions
 for particles and anti-particles. 
Evaluating the derivative with respect to $k^2$ on the right  hand side,
       we obtain 
\be
\label{eq:fermi-2}
\frac{\pd \Omega^F}{\pd k^2}&=&-\frac{N_cN_f}{32\pi^2}k^4
        \int\limits_0^{\infty}\!dq\,\biggl[\frac{1}{E_{q,k}^3}(1-n(E_{q,k})
        -\bar{n}(E_{q,k}))   \nonumber \\   
          & & - \f{1}{TE_{q,k}^2}\{n(E_{q,k})(1-n(E_{q,k}))
          +\bar{n}(E_{q,k})(1-\bar{n}(E_{q,k}))\}\biggl].
\ee
We notice that the distribution functions explicitly show 
how the temperature and the baryonic density modify the  
$T=0$ and $\rho_B=0$ result:  The 
right hand side of the evolution equations in vacuum is 
diminished (1) by the 
Pauli blocking effect as seen in the first bracket, and (2)
by the thermally excited states 
as seen in the second bracket.
Similarly, the bosonic part at $T\not=0$ is evaluated as shown in appendix A,
where $\pd \Omega^B/\pd k^2$ is expressed in terms of the Bose-Einstein
distribution function. 

Although formula~({\ref{eq:fermi-2}) is generally valid for $T\not=0$
and $\rho_B\not=0$, it is not useful when the temperature is set to zero
due to the factor $T^{-1}$ in the second bracket. The zero temperature
formula is obtained easily from eq.~(\ref{eq:fermi-1})
making the substitutions
 $n(E_{q,k})\rightarrow \theta(\mu-E_{q,k})$ and
 $\bar{n}(E_{q,k})\rightarrow 0$:
\be
 \label{eq:flow-fermion}
 \left.\partl{\Omega^F}{k}\right|_{\mu} =
         -\f{N_cN_f}{8\pi^2} \f{k^5}{k^2\!+\!g^2\phi^2} 
        \left[1\!-\!\frac{\mu}{\sqrt{\mu^2-k^2-g^2\phi^2}}
\Theta(\mu\!-\!\sqrt{k^2+g^2\phi^2}) \right] .
\ee
Here again we see  how the evolution equation at finite baryon density is
modified by Pauli blocking. In appendix \ref{sec:connection} we show how the
integral of the finite density part of equation (\ref{eq:flow-fermion}) gives
the mean field result, which will be presented in section \ref{sec:coarse-grained-pot}. 

For the charge neutral system (zero bosonic
chemical potential) at $T=0$ the bosonic term is the same as in vacuum.
Adding both contributions, we have the zero temperature flow equation for finite
baryon density as follows: 
\be
  \left.\partl{\Omega}{k}\right|_{\mu} &=& 
        \displaystyle \f{k^5}{32\pi^2} \left\{
        \f3{k^2\!+\!2\Omega^{\prime}}\!+\!
        \f1{k^2\!+\!2\Omega^{\prime}\!+\!4\Omega^{\prime\prime}\phi^2}
           \right.\\
        && \displaystyle \left.
        \!-\f{4N_c N_f}{k^2\!+\!g^2\phi^2} \left( 1 \!-\!
          \f{\mu}{\sqrt{\mu^2\!-\!k^2\!-\!g^2\phi^2}}
                \Theta(\mu\!-\!\sqrt{k^2\!+\!g^2\phi^2})
        \right) \right\}. \nonumber
\ee
We stress that such a compact form is obtained for the flow equation 
in the heat-kernel method; the Pauli blocking effect due to the presence of
the Fermi sea is explicitly represented by a theta-function.
This is certainly an advantage of the method. 
However, the presence of the theta-function makes
numerical evaluations difficult because its derivatives 
produce singular terms
which are not easy to control in the numerical analysis.

As an  attempt to circumvent this difficulty, we let the
chemical potential run
during the evolution; so the
chemical potential will be a function of the scalar field and the momentum
scale $\mu=\mu(\phi,k)$, as proposed by Shankar in reference~\cite{Shankar}.
In fact, it turns out that no singularities of the fermion-terms appear in the
evolution with a running $\mu$. In the functional integral for
$\Omega$ we explicitly insert a $k$-dependent chemical potential.

The method is best explained starting from $k=0$. At the end of the
evolution the transition to the free energy density can be made via a a
Legendre transformation.
\be
  F(k\!=\!0,\phi^2) &=& \Omega(k\!=\!0,\phi^2) +  \rho
        \mu(k\!=\!0,\phi^2), 
\ee
where $\mu(k=0)$ has to be eliminated from $\Omega(k\!=\!0,\phi^2)$
via the equation 
\be
  \label{dens}
  \partl{\Omega(k=0)}{\mu(k=0)} = -\rho.
\ee
Now we make a small change of infrared cut-off scale $k$
and adjust $\mu(k)$ in such a way that the density remains constant:
\be
  \label{dens}
  \partl{\Omega}{\mu(k)} = -\partl{\Omega^F}{\mu(k)} = -\rho.
\ee
The same procedure is repeated at each step of $k$.
The explicit evaluation of the l.h.s. at a finite evolution scale $k$
using~(\ref{omferm}) yields
\be
  \label{dens-2}
  \rho = \frac{2N_c}{3\pi^2}
  \left(x_F^3+{3\over 2}k^2x_F+{3\over 8}{k^4\over x_F} \right),
\ee
with
\be
  \label{x}
  x_F=\sqrt{\mu^2-k^2-g^2\phi^2}.
\ee

An analysis of  the solutions $\mu(x_F)$ from eq.~(\ref{dens-2}) shows
that 
at $k=0$ this equation has a unique solution
$\mu=\sqrt{g^2\phi^2+k_F^2}$, with the Fermi momentum $k_F$ defined by 
$\rho=2N_ck_F^3/3\pi^2$. 
For $k\neq0$ it is advantageous to discuss the solutions 
in terms of $z=x_F/k$. Equation (\ref{dens-2}) then  reads
\be
\label{rho}
  \rho = \f{2 N_c}{3\pi^2} k^3\cdot g(z),
\ee
with
\be
\label{eq:constraint}
g(z)= z^3+\f{3}{2}z+\f{3}{8z}.
\ee
{\begin{figure}[t]
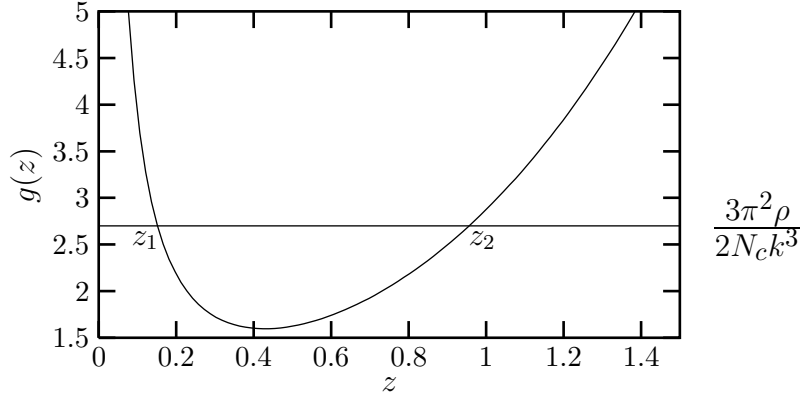

    \centerline{\input figure1.tex}
    \caption{The function $g(z)$ with $z=x_F/k$, which allows to find the
      solutions of the constraint 
      eq.~(\protect\ref{eq:constraint}) at a fixed density $\rho$.}
    \label{fig:constraint} 
\end{figure}}
The behavior of $g(z)$ is shown in Figure~\ref{fig:constraint}; 
$g(z)$ has the minimum  
\begin{equation}
 g_{min}=\frac{1}{2}\frac{\sqrt{3}+1}{\sqrt{\sqrt{3}-1}} 
\end{equation}
at $z=\frac{1}{2}\sqrt{\sqrt{3}-1}\equiv z_m$.
At finite 
evolution scale $k$ the line of 
constant $\frac{3 \pi^2\rho}{2 N_ck^3}$ will in general
cut the constraint function
at two points $z_1$ and $z_2$. It is necessary for the equation to have real
solutions that 
\be
k\le \left(\f{3\pi^2\rho}{N_c} \f{\sqrt{\!\sqrt{3}\!-\!1}}%
        {\sqrt{3}+1} \right)^{1/3}\equiv k_{\rm th}.
\ee 

At $k=k_{\rm th}$ the line of 
constant $\frac{3 \pi^2\rho}{2 N_ck^3}$
is tangential
to $g(z)$ at $z=z_m$.
Corresponding to the two solutions, the chemical potential has
two different values at fixed density:
\be
\mu_i(k)=\sqrt{k^2z_i(k)^2 +k^2 +g^2\phi^2} \quad (i=1, 2).
\ee
The derivatives $\pd \mu_i/\pd k$ 
are calculated by demanding that the baryon density
found at $k=0$ remains the same
{\em independently} of the evolution scale $k$. Thus the $k$ derivative
of the l.h.s of eq.~(\ref{dens-2}) is zero, leading to
\be
\frac{\pd \mu_i}{\pd k}= -{1\over \mu_i}\frac{k}{8z_i^4+4z_i^2-1}.
\ee
Applying the chain rule one can calculate the flow equation
for the thermodynamic potential with running chemical potential,
\begin{equation}
\label{eq:chainrule}
 \left. \frac{\partial \Omega}{\partial k}\right|_{\mu(k)}
 = \left. \frac{\partial \Omega}{\partial k}\right|_{\mu=const.} 
+ \frac{\partial \Omega}{\partial \mu}\frac{\partial \mu(k)}{\partial k}
 = \left. \frac{\partial \Omega}{\partial k}\right|_{\mu=const.} 
  -\rho \frac{\partial \mu(k)}{\partial k} ,
\end{equation}
or rearranging this formula for each k we get the
equivalent free energy at each $k$:
\be
\left.\partl{F}{k}\right|_{\rho_{{}_B}} =
\left.\partl{(\Omega+\rho\mu)}{k}\right|_{\mu(k)} =
\left.\partl{\Omega}{k}\right|_{\mu=const.} \,.
\ee
We can eliminate   the chemical
potential  in the flow-equation
at each  $k$ via a $\delta$-function,
\be
 1 = \int d \! f \; \delta\left[f(z)\right],
\ee
with 
\be
f(z)=\f{2 N_c}{3\pi^2} k^3g(z) -\rho \, .
\ee
The evolution equation (\ref{eq:chainrule}) with running
chemical potential
has two terms $i=1,2$ corresponding to the two roots of the
constraint equation:
\be
  \left. \frac{\partial \Omega}{\partial k}\right|_{\mu(k)} 
& = &  \int d \! f \; \delta\left[f(z)\right] \left[ \left. \frac{\partial
  \Omega}{\partial k}\right|_{\mu=const.} + \frac{\partial \Omega}
{\partial \mu}\frac{\partial
  \mu(k)}{\partial k} \right] \\
& = &  \sum_{i=1,2}\int  \frac{\partial f}{\partial z}  d \! z \; 
  \left| \frac{\partial f}{\partial z} \right|^{-1}
  \delta (z\!-\!z_i) \left[ \left. \frac{\partial
  \Omega}{\partial k}\right|_{\mu=const.} -  \rho \frac{\partial \mu(k)}
      {\partial k} \right]. 
\ee

Note that at the two zeros of the function $f$, $z_1$ and $z_2$, the
Jacobians have opposite signs (see figure~\ref{fig:constraint}) leading
to some cancellation. Thus  we arrive at the final flow equation 
\be
\label{eq:fflow}
\left. \partl{\Omega}{k}\right|_{\mu(k)} &=& 
\displaystyle \f{k^5}{32\pi^2} \left\{
        \f3{k^2\!+\!2\Omega^{\prime}}\!+\!
        \f1{k^2\!+\!2\Omega^{\prime}\!+\!4\Omega^{\prime\prime}\phi^2}
        \!-\!\f{8N_c}{k^2\!+\!g^2\phi^2} \right\} \\
   &+& \displaystyle \sum_{i=1,2} (-1)^i \left\{ \f{N_c}{4\pi^2} 
        \f{k^4}{k^2\!+\!g^2\phi^2} \f{\mu_i}{z_i} 
        \!+\!\f{\rho}{\mu_i}\ \f{k}{8z_i^4+4z_i^2-1} \right\} 
\Theta(k_{\rm th}-k).
        \nonumber
\ee
The reason of the presence of  the theta-function in (\ref{eq:fflow})
 is apparent from~(\ref{omferm}). Note that the term $1/(8z_i^4\!+\!4z^2_i
\!-\!1)$ is singular at $k\!=\!k_{th}$ but can be integrated analytically. 
\begin{figure}[t!]
  \begin{center}
    \centerline{\input{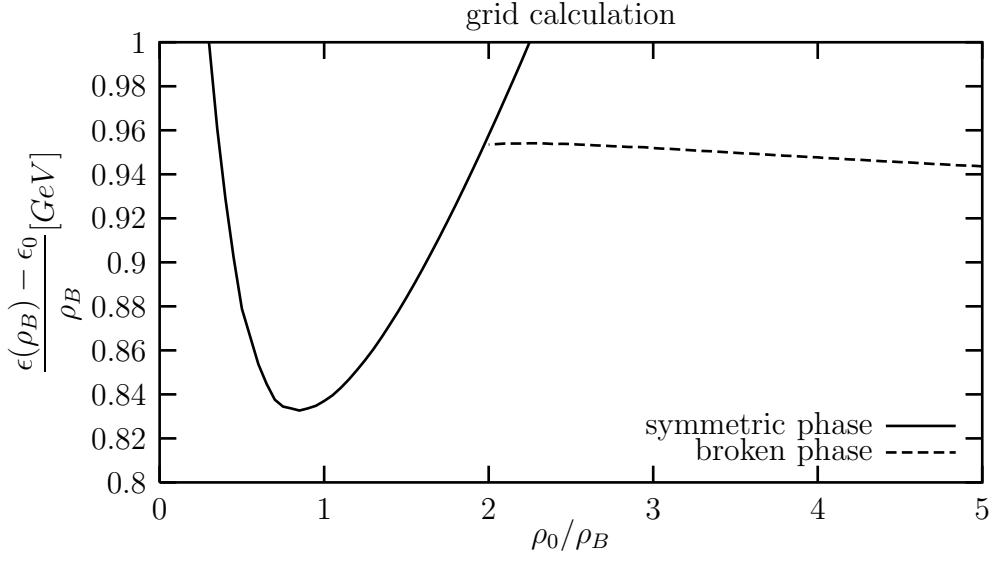}} 
    \footnotesize
   \parbox{12cm}{
   \caption{\small Normalized energy per baryon plotted against inverse density
     $\rho_B^{-1}$, normalized to normal nuclear density $\rho_0 = 0.16$
   fm$^{-3}$, in the broken phase (dashed line) and in the symmetric phase
   (solid line). The evolution eq.~(\protect\ref{eq:fflow}) is solved on
   the grid.  The energy of the broken phase at zero density is $\epsilon_0$.}
   \label{fig:grid-en-plot}
   }
  \end{center}
\end{figure}
\begin{figure}[h!]
  \begin{center}
     \centerline{\input{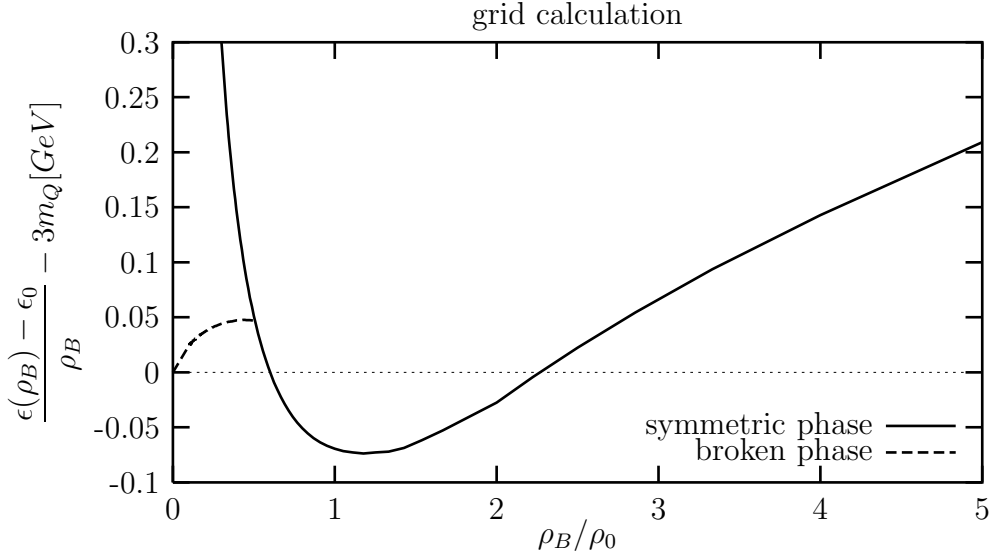}}
    \footnotesize
   \parbox{12cm}{
   \caption{\small The binding-energy per baryon 
    calculated from the evolution equation on the grid is
    shown as a function of $ \rho_B $ normalized to
     normal nuclear density $\rho_0 $. 
     \label{fig:grid-binding}
   }
 }
  \end{center}
\end{figure}

Equation~(\ref{eq:fflow}) is the evolution equation with running
chemical potential $\mu(k)$
at zero temperature for the thermodynamic potential density. In
order to follow the minimum of the order parameter at fixed density we
need the free energy density.  Recall that at $k=0$ there is only one
solution to the constraint equation~(\ref{dens-2}) which is related to
$z_2$, therefore we can calculate the free energy density $F$ at $k=0$
from the thermodynamic potential density $\Omega(k=0)$ and the chemical
potential $\mu(k=0)$:
 \be
  F(k\!=\!0,\phi^2) &=& \Omega(k\!=\!0,\phi^2) +  \rho
        \mu(k\!=\!0,\phi^2) \\
  \mu(k=0,\phi^2)&=& \sqrt{k_F^2+g^2\phi^2} \nonumber
\ee

Due to the theta term in eq.~(\ref{eq:fflow}) the $k$ evolution of
$\Omega(k)$ and $F(k)$ is the same till $k_{th}$ and does not feel the
baryon density.  Below $k_{th}$ the density effects set in with the
contributions of a shell of fermions. During the course of evolution the
outside radius of the shell increases and the inside radius diminishes until the
longest wavelengths in the fermi sea are integrated.  The effect arising
from the fermi sea is classical, hence ${\cal O}(\hbar^0)$ and dominates
the RG flow equation. For large enough densities $\rho\
\raisebox{-3pt}{$\stackrel{>}{{\scriptsize\sim}}$} \ 0.45 \rho_0$, we
have $k_{th}>k_{inst}$ and eq.~(\ref{eveq}) is replaced by the density driven
evolution without unstable boson terms.

We solved eq.~(\ref{eq:fflow}) numerically on a grid: the full potential
density $\Omega$, is discretized as a function of $\phi^2$ on a grid of
hundred points between $0 < \phi^2 < 0.05 \; {\rm GeV}^2$. The resulting
hundred differential equations are solved with a Runge Kutta method.

As we discussed earlier at the end of section~\ref{sec:evol-lin-sig} the meson
terms in the flow equation~(\ref{eq:fflow}) develop singularities due to
the instability of the effective potential. This behavior is well known in the
literature~\cite{Rho,Bentz}, the mesonic effective potential in the
one loop approximation generates tachyonic meson masses. The
renormalization group scale $k^2$ avoids these poles for some part of
the evolution, but cannot get rid of them totally. The singularities
in the boson denominators of the evolution
equation appear at $k_{inst}^2+2\Omega^\prime=0$, 
indicating the disappearance of the Gaussian fix
point~\cite{Janos}. As our studies show the inclusion of the
non-Gaussian fixed point does not change the position of the minimum
considerably at zero temperature. 
%
%
Hence in this paper we neglect the singular
mesonic contributions beyond $k=k_{inst}$ in the evolution equation
leaving the more exact solution to a future work. 
The results
of the grid calculation are shown in figures \ref{fig:grid-en-plot} and
\ref{fig:grid-binding}. We will discuss them extensively
in section \ref{sec:meangrid} and
compare them with the mean field calculations presented in the next section.

\section{Coarse grained potential in mean field approximation}
\label{sec:coarse-grained-pot}

Since the numerical solution of the RG flow equations has 
difficulties at small momentum scale $k$,  we discuss 
another approximation to the low momentum region in this section.
We evolve the
vacuum theory from the ultraviolet scale $k=k_{uv}$ down to $k = k_{F}$
corresponding to normal nuclear matter density $\rho_B=\rho_0$.  Thereby
we obtain a coarse grained potential which is appropriate for the
dynamics at these low momenta. This potential contains the vacuum loop
effects of the quarks and bosons integrated out up to this
scale. Then we solve this coarse grained linear sigma model in
mean field approximation.

\begin{figure}[h]
  \begin{center}
     \centerline{\input{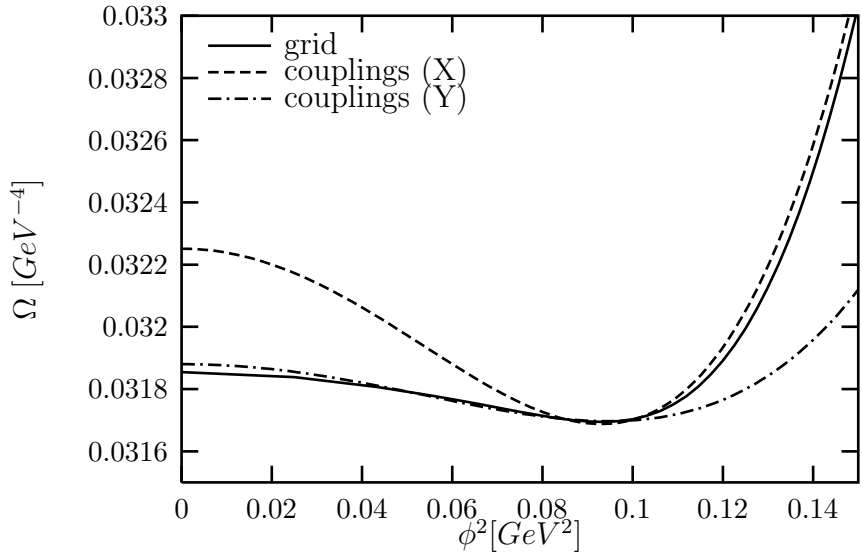}}
    \footnotesize
   \parbox{12cm}{
   \caption{\small The potential $\Omega(\phi^2)$ at momentum scale $k =
     k_{th}$, from the grid calculation (solid line) and the fitted
     curves with mean field parameters $X$ (dashes) and $Y$
     (dashed-dotted line). See table~\protect\ref{tab:couplings}.
     \label{fig:pot}
   }
 }
  \end{center}
\end{figure}

To keep the mean field approximation transparent we approximate the coarse
grained potential with the original fourth order form of equation~(\ref{eq:pot})
with renormalized parameters.  These parameters depend on the range in
$\phi^2$ where the fit to the coarse grained potential is done. We have
chosen two sets: a fit on a 
wider region in $\phi^2$ (X) and a narrower one (Y). The resulting
parameters are summarized in Table~\ref{tab:couplings}. The coarse grained
potential and the two fits are shown in figure \ref{fig:pot}.  The negative
values of the mass squared indicate that at the scale $k=k_F$ we already
entered the broken phase, with a nonvanishing expectation value of
the $\sigma$ field, $\bar\sigma$. A mean field solution with these
potentials is straightforward by evaluating the Hamiltonian $H$, in the
presence of the Fermi sea and then minimizing $H$ with respect to
$\bar\sigma$.  Since we have already taken into account quantum
fluctuations from the Dirac sea and the bosons in the evolution, we do
not need to consider these pure quantum fluctuations any longer.  Their
effects are assumed to be of higher scale than $k_F$ and therefore they
are integrated up in the coarse grained potential. Of course, we are
missing typical quantum many body fluctuations not included in the
mean field approximation of the many body system.

\renewcommand{\arraystretch}{1.2}
\begin{table}[htb]
  \begin{center}
 \begin{tabular}{c|c|c|c|c|c|c|c}
   Type & Fit range $\phi^2$ & $m^2_{MF}$ & 
                $\lambda_{MF}$ & $ \bar
   \sigma_0$ & $3g \bar\sigma_0 $ & $B^{1/4}$ & $m_\sigma$\\
\hline
  $X$ & $\left[ 0,0.05 \right]$ & -0.260 & 30.0 & 0.0940& 0.902 &0.15554 & 0.720 \\
  $Y$ & $\left[ 0,0.01 \right]$ & -0.082 & 9.18 & 0.0945& 0.906 &0.11633 & 0.405 \\
  grid&-&-&-&                                     0.0944& 0.906 & - & - \\
\end{tabular}
    \caption{Effective meson potentials used. The fit range and the mass
    parameter $m^2_{MF}$ are in {\rm GeV}$^2$, the $\bar\sigma_0$, 
    $B^{1/4}$ and the mass of the $\sigma$ meson are in {\rm GeV}.}
    \label{tab:couplings}
  \end{center}
\end{table}

The mean field solution of the resulting $\phi^4$ theory 
is standard. The
Hamiltonian reads
\begin{equation}
  \label{eq:hamilton-fcn}
  H = \int\!d^3x\, \left[ q^\dagger \left( \vec \alpha \vec p  +  g
  \beta \bar
  \sigma \right) q + \frac{m^2_{MF}}{2}\bar \sigma^2 
  + \frac{\lambda_{MF}}{4} \bar \sigma^4 \right].
\end{equation}
Using a plane-wave basis for the quarks 
one can rewrite the Hamiltonian as an integral over momenta.
In the ground-state, $T = 0$, the Fermi-sphere is filled from the bottom
up to the Fermi-momentum $k_F$. The energy density is evaluated to be
\begin{eqnarray}
  \label{eq:energie}
  \epsilon = \frac{E}{V} &=& \frac{4 N_c}{(2 \pi)^3} 
             \int\limits_0^{k_F} \!d^3k\; \sqrt{\vec k^2 + g^2\bar \sigma^2} +
             \frac{1}{2} m^2_{MF}\bar \sigma^2 + \frac{\lambda_{MF}}{4} \bar \sigma^4\\
 &=& \frac{N_c}{4\pi^2} \left(2k_F \sqrt{k_F^2+g^2\bar \sigma^2}^3
                      - g^2\bar \sigma^2 \, k_F\sqrt{k_F^2+g^2\bar \sigma^2} \right. \nonumber \\
                 & & \left.    - g^4\bar \sigma^4  
                      \log{\frac{\sqrt{k_F^2+g^2\bar \sigma^2}+k_F}{g\bar
             \sigma}} \right) + \frac{1}{2} m^2_{MF}\bar \sigma^2 + \frac{\lambda_{MF}}{4} \bar \sigma^4.
       \nonumber
\end{eqnarray} 
The  quark Fermi-momentum is fixed by the quark-density $\rho=3\rho_B$,
\begin{equation}
  \label{eq:fermi-impuls}
  k_F = \sqrt[3]{\frac{3\pi^2 \rho}{2 N_C}}.
\end{equation}

The mean $\sigma$-field configuration is calculated by minimizing
(\ref{eq:energie}) with respect to $\bar \sigma$. One ends up with a
self-consistent equation for the mean field $\bar \sigma$, 
\begin{equation}
  \label{eq:energie-minimum}
  \left( \frac{\partial \epsilon}{\partial \bar \sigma} \right)_V 
 = \frac{4 N_c}{(2 \pi)^3}  \int\limits_0^{k_F} \!d^3k\,
\frac{g^2 \bar \sigma}{\sqrt{\vec k^2 + g^2\bar \sigma^2}} + m^2_{MF}\bar \sigma +
 \lambda_{MF} \bar \sigma^3 \stackrel{!}{=} 0.
\end{equation}
There is always the trivial solution $\bar\sigma=0$ corresponding to
the symmetric phase.  At zero density the non trivial solution $\bar
\sigma_0 = \sqrt{-m^2/\lambda}$ has a lower energy and represents the
spontaneous broken phase.  At higher density, the chirally symmetric
phase has lower energy. The energy density in the broken phase relative
to the symmetric one is
\begin{equation}
  \label{eq:energie2}
  \epsilon_0 \equiv \epsilon_{br}(\rho_B=0) =  -\frac{m^4}{4\lambda} = -B,
\end{equation} 
whereby we define $B$ as a kind of bag constant.  It gives the amount of
energy density by which the partonic vacuum lies above the constituent
quark vacuum.  We estimate the sensitivity of the calculation to the
input coarse grained potential by comparing the minima of the energy per
baryon in both phases. The massless partonic (symmetric) phase has an
energy density
\begin{equation}
\epsilon_{sym} = \frac94 \left(\frac{9\pi^2}{2 N_c}\right)^{1/3}
        \rho_B^{4/3}
\end{equation}
with a minimum of the normalized energy per baryon
\begin{equation}
\label{eq:part-phase}
\left. \frac{\epsilon_{sym}(\rho_B)-\epsilon_0}{\rho_B}\right|_{min}
=3 \left(\frac{3\pi^2}{2 N_c}\right)^{1/4} 
        \left(\frac{m^4}{\lambda}\right)^{1/4}
\end{equation}
at
\begin{equation}
  \rho_{B,sym}^{min} = \frac13 \left(\frac{2 N_c}{3\pi^2}\right)^{1/4} 
        \left(\frac{m^4}{\lambda}\right)^{3/4} \,.
\end{equation}
In the constituent quark phase the asymptotic value of the energy per
baryon is three times the quark mass:
\begin{equation}
\label{eq:const-phase}
\left. \frac{\epsilon_{br}(\rho)-\epsilon_0}{\rho_B}\right|_{min}=3 g \bar
\sigma_0 \,.
\end{equation}
Therefore in order to have a stable broken phase at low density
the following condition is necessary (but not sufficient),
\begin{equation}
\label{eq:ineq}
\left(\frac{3\pi^2\lambda}{2 N_c}\right)^{1/4} > g \,.
\end{equation}
We note, that with $g$=3.23 this condition is fulfilled for
parameterization $(X)$ but not for $(Y)$. As a result the latter does not
have a stable homogeneous phase with broken chiral symmetry.

In the following section we present our numerical results.

\section{Results in mean field approximation}
\label{sec:result-lin-sig}
In this section we discuss the results of the mean field calculations, shown in the figures
\ref{fig:1}--\ref{fig:2}, before we compare them with the grid calculation in
the next section.
In figure~\ref{fig:1} we present the energy  per baryon as a function of the
inverse density relative to the energy of the broken phase
at zero baryon density. In the low
density limit, the energy per baryon of the broken phase slowly approaches 
three times the constituent quark mass $3m_Q = 3g \bar \sigma_0$.  
This is the preferred state at low densities,
where we have a noninteracting dilute system of constituent quarks.
Since our coupling constant $g$, is fixed from
the beginning to yield a constituent quark mass of $300 \; {\rm MeV}$, the vacuum
mass of the ``nucleon'' is smaller than $938 \; {\rm MeV}$ in both cases (X)
and (Y) (cf. table \ref{tab:couplings}). In the chiral limit which we pursue
here the nucleon is lighter than the real one. 
\begin{figure}[thb]
  \begin{center}
     \centerline{\input{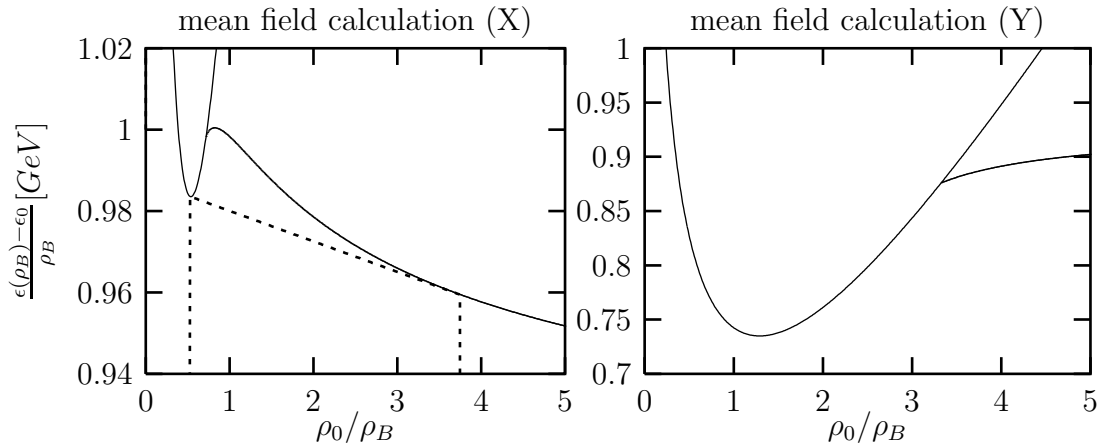} }
    \footnotesize
   \parbox{12cm}{
   \caption{\small The energy per baryon plotted against 
        $ \rho_B^{-1}$ in units of normal nuclear density
        $\rho_0=0.16 \; fm^{-3}$ for mean field parameterization $X$
   (left) and $Y$ (right). Note the different energy scales in the
        figures.         
        The dashed line in the left plot represents the
     Maxwell-construction which determines the region of coexistence between
     broken and chiral symmetric phase.}
   \label{fig:1}
   }
  \end{center}
\end{figure}
In the upper left corner of both plots the parabola
represents the partonic phase.

For the couplings (X) condition (\ref{eq:ineq}) is fulfilled, thus 
the phase diagram may be obtained using
the Maxwell construction (dashed line):

\begin{figure}[t]
  \begin{center}
     \centerline{\input{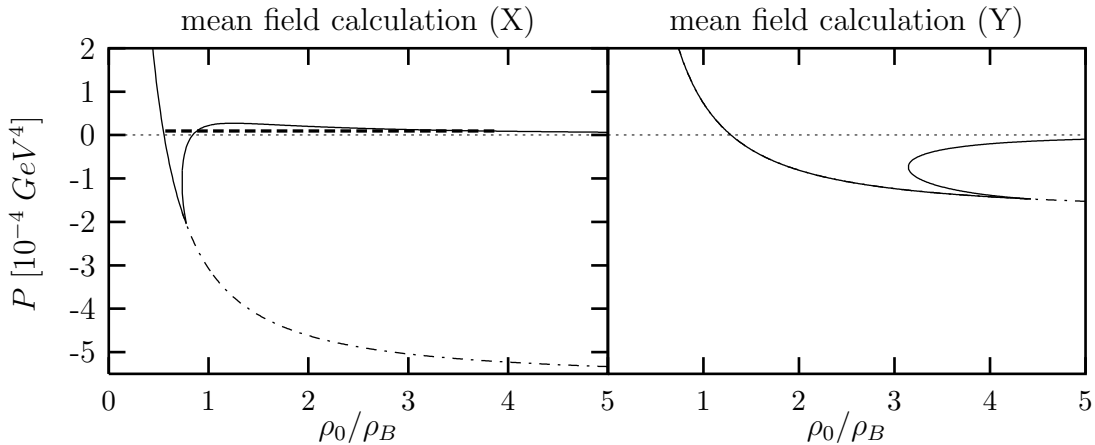}}
    \footnotesize
   \parbox{12cm}{
   \caption{\small The pressure as a function of inverse baryon density, 
    calculated for mean field parameterizations $X$ (left) and $Y$
    (right). The thick dashed line in the left plot is the Maxwell-construction. In
    the case $Y$ (right) no Maxwell-construction exists. Note that the negative
    values of the pressure are solutions of the equations but have now physical
    interpretation. In both cases the pressure for the symmetric phase ($\bar \sigma = 0$)
    asymptotically approaches the bag pressure $B$ (cf. table
    \ref{tab:couplings}) for small densities (dashed-dotted line).}
     \label{fig:4}
   }
  \end{center}
\end{figure}

For a first order phase transition the equilibrium-condition for a given
temperature (in our case $T=0$) and pressure is
\begin{equation}
  G(T,P,N) = \mbox{min}\,.
\end{equation}
In the region of coexistence between
the two phases (I/partonic) and (II/con\-sti\-tuent quark), the 
temperature, pressure and chemical potential have
to be equal to each other. Hence
\begin{eqnarray}
  \mu^{I,II}_B = \frac{1}{N_B} G(T,P,N_B) 
            = \frac{1}{N_B} (F+PV)
            = \frac{F}{N_B} + P \frac{1}{\rho_B},
\end{eqnarray}
where the pressure $ P $ is constant in the coexistence region. 
At zero temperature the free energy per baryon $F/N_B$ equals the
energy per baryon $E/N=\varepsilon/\rho_B$ and we  
can read off from the
tangent construction the energy per baryon 
as a function of the inverse baryon density
in the mixed phase:
\begin{equation}
  \label{eq:pressure}
  \frac{E}{N_B} = -P\frac{1}{\rho_B} + \mu_B.
\end{equation}
The slope is the negative pressure and the intercept with the vertical
axis is the baryon chemical potential at the phase transition. 
The dashed line in figure~\ref{fig:1} connects the low density
constituent phase with the high density partonic one. The phase
transition takes place between $0.27$ and $1.90$ times normal nuclear
density. In between the two phases coexist.

The couplings (Y) do not fulfill condition (\ref{eq:ineq}), e.g. we have no
stable broken phase. Hence there is no Maxwell construction in this case.  

\begin{figure}[t]
  \begin{center}
     \centerline{\input{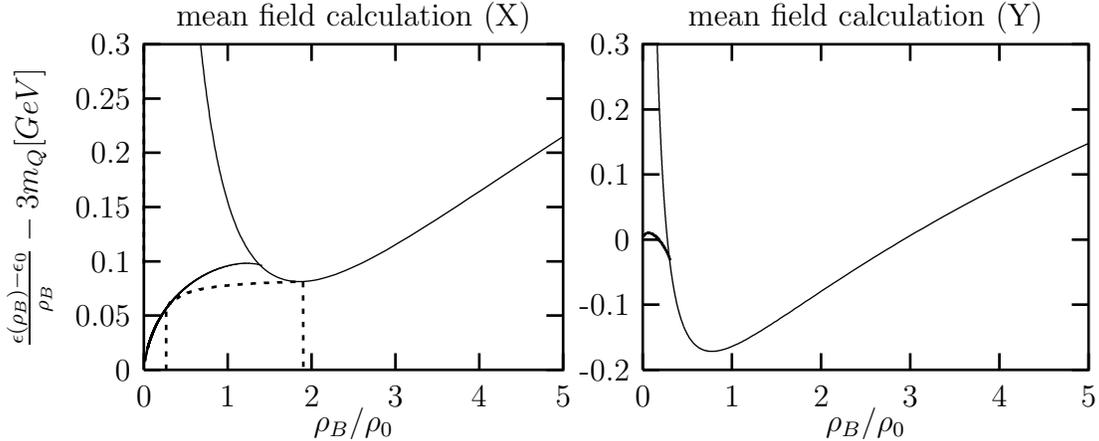}}
    \footnotesize
   \parbox{12cm}{
   \caption{\small Binding energy per baryon is plotted against 
    $ \rho_B $ in units of 
     normal nuclear density $\rho_0$, for 
     the couplings $X$ (left panel) and $Y$ (right panel).}
     \label{fig:2}
   }
  \end{center}
\end{figure}

For both parameterizations (X) and (Y) we calculate the pressure 
from equation (\ref{eq:pressure}) . The result is shown in figure \ref{fig:4}.
It shows clearly the stable broken phase for parameterization (X). For
parameterization (Y) there is no stable broken phase. At low densities the pressure is always
negative, thus it is more advantageous to pack the quarks into droplets
of massless quarks with nonperturbative vacuum between them than to have
loosely bound constituent quarks associated with nucleons.

In figure~\ref{fig:2} we present the binding energy per baryon
subtracting the ``baryon  mass'' $3 m_Q$, from the energy per baryon.
Note that in the case (X) the system is unbound, the energy of the partonic
phase has a minimum at $\sim 1.8$ times normal nuclear
density. This minimum lies on the edge of the coexistence region.
In the case of couplings (Y) the system is strongly bound, the minimum of the
partonic phase is $\rho_B = 0.78 \, \rho_0$.  

The difference between the two cases $(X)$ and $(Y)$
demonstrates that the form of the coarse grained potential $\Omega$
influences 
the physics at low momentum scale drastically. In the 
mean field approximation it is mainly the sigma mass
$m_{\sigma} ^2=2 \lambda \bar \sigma_0^2$ which 
determines the amount of attraction i.e. binding or unbinding
of baryonic matter. Recall the attraction in
nonrelativistic Hartree approximation varies as $-\frac{g^2 \rho_s^2}
{2 m_{\sigma}^2}$.
Potential (X) with a high $\sigma$-mass of $0.728 $ GeV is less
attractive than the potential (Y) with a $\sigma$-mass of $0.404
$ GeV. The meson--meson interaction term $\lambda$, also fixes the
structure of the intermediate density region. A large $\lambda$ gives a
mixed phase as produced by potential (X) cf. eq.~(\ref{eq:ineq}). The grid
calculation shares the low field strength region with the potential (Y),
and has the mass of the $\sigma$ meson in between of the ones from the
potentials (X) and (Y) (see figure \ref{fig:pot}).  Therefore we expect
that the equation of state on the grid lies between the extremes
determined by the potentials (X) and (Y).

\section{Comparison of mean field results with grid calculations}
\label{sec:meangrid}

In this section we compare the results obtained from the mean field
approximations with the ones from the grid calculation.
The main results are shown in figures~\ref{fig:comparrison1} and~\ref{fig:comparrison2}. 

\begin{figure}[t]
  \begin{center}
     \centerline{\input{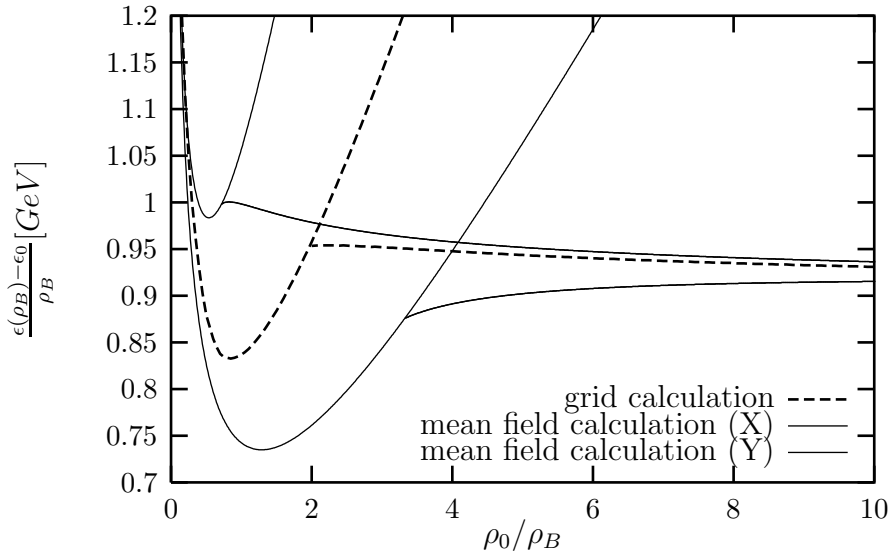}}
    \footnotesize
   \parbox{12cm}{
   \caption{\small Normalized energy per baryon, calculated on the grid
     (dashed line) and in the mean field calculations 
     with couplings $X$ (upper solid curve) and $Y$ (lower solid curve),
     is  shown as a function of inverse density $ \rho_B^{-1}$,
     normalized to normal nuclear density.  
     \label{fig:comparrison1}
   }
 }
  \end{center}
\end{figure}
\begin{figure}[t]
  \begin{center}
     \centerline{\input{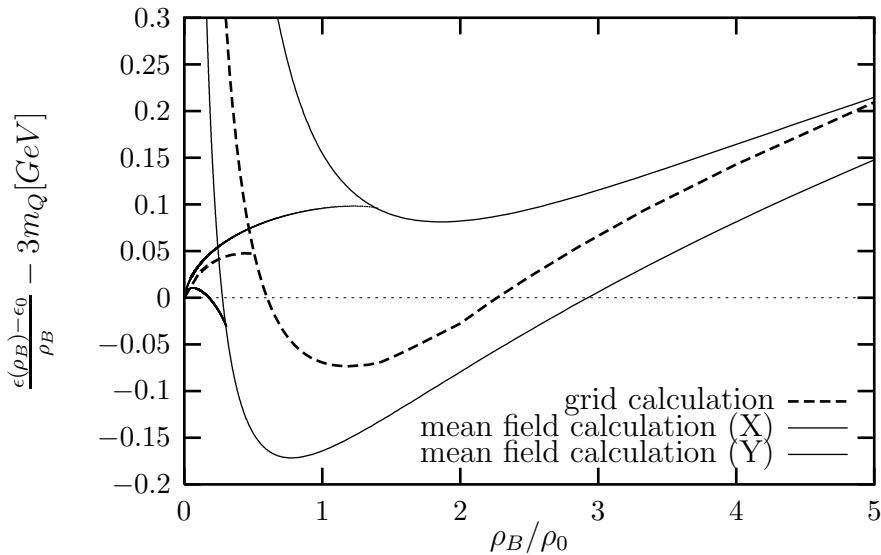}}
     \footnotesize
   \parbox{12cm}{
   \caption{\small Binding energy per baryon, calculated on the grid and
     in the mean field approximation with potentials $X$ (upper curve)
     and $Y$ (lower curve), is  shown as a function of $ \rho_B $
     normalized to normal nuclear density.  
     \label{fig:comparrison2}
   }
 }
  \end{center}
\end{figure}

In figure~\ref{fig:comparrison1} we present the energy per baryon obtained in the
(X) parameterization of the mean field (upper curve), the (Y) parameterization
(lower curve) and the grid calculations
(dashed line). The three curves show a very similar behavior. At low density
the energy per particle is approaching the same limit of three times the
mass of the constituent quarks, as we discussed in the previous
section. The nucleon mass is $\sim 900$ {\rm MeV} in both cases. With
increasing density, one arrives at the point where no broken phase is
supported any longer, i.e. 
the mean field 
eq.~(\ref{eq:energie-minimum}) does not have a nontrivial $\bar\sigma$
solution and only the massless partonic phase exists.  For the
coarse grained couplings (Y), this happens at $\rho_B=0.32\, \rho_0$, 
while for the grid calculation at $\rho_B=0.56\, \rho_0$. The parameterization
(X) leads to highest transition point at $\rho_B = 1.32  \,\rho_0 $. In all
three cases the order parameter drops to zero, i.e. we have a first order
phase transition.

Since the minimum of the energy per particle in the partonic phase is
below the minimum of the broken phase, no Maxwell-construction is
possible and the broken phase is not stable even at lower
densities for both the grid calculation and model (Y).
This phase 
starts from zero density and persists until $\rho_B=1.17\, \rho_0$ for
the grid and until $\rho_B=0.78\, \rho_0$ for 
the mean field calculation (Y). At these
densities the droplets fill up the whole volume. They are bound by
$\approx 74 \,{\rm MeV}$ (grid) and $\approx 172 \,{\rm MeV}$ (mean field),
respectively,  per baryon number.
If ones compresses the system above this density the fermi
pressure pushes the equation of state higher up in energy. 

Contrary to these cases  there exists
a Maxwell construction for potential (X)
as we have seen in the previous section. The region of
coexistence between the stable constituent quark phase and the partonic phase
ranges from $0.27\,\rho_0$ to $1.90 \, \rho_0$.

In figure~\ref{fig:comparrison2} we show the equation of state for the three calculations.
The coarse-grained couplings (Y) as well as the
grid-calculation yield  bound systems. However, in both cases the
binding is too strong, $74$ {\rm MeV} per baryon in the  grid and $172$
{\rm MeV} per baryon in the mean field calculation with model
(Y). The saturation densities
lie near normal nuclear density, but baryonic matter is
already in a chirally symmetric phase.

\section{Summary and Conclusion}
\label{sec:conclusion}
We have calculated a coarse grained effective potential
from renormalization group flow equations in a quark model with
explicit meson fields. At the fermi momentum scale we continue
the evolution including Pauli blocking. To avoid meson 
instabilities we switched off the meson loop terms at $k_{th} \approx k_F$.
The resulting quark matter overbinds and is in a chiral
symmetric phase. We presented two mean field approximations to the full grid
calculation with different sets of couplings. The grid solution of the
flow equation, with the evolution of the meson effective potential 
omitted below $k_{th}$, lies in between 
the result of these two mean field approximations.

The many body physics found here is very similar to the \emph{NJL} model
where one finds a first order phase transition with either a mixed phase
or a droplet phase~\cite{Buballa}. Also the instanton induced quark interaction of
ref.~\cite{DIQUARK1} produces a droplet phase of partonic quarks.  In
the effective potential the determining equation is the relation between
the energy per quark in the partonic phase and the constituent quark
mass (cf. eqs.~(\ref{eq:part-phase},\ref{eq:const-phase})).

The  shape of the droplets in the low density phase 
have to be determined from an
independent calculation including surface effects.
The  bag produced by the linear sigma model 
does not confine. It has a finite height, outside the massless
quarks acquire a constituent quark mass.
Otherwise the solution of the evolution equation is similar to a
\emph{MIT} bag
type solution with massless quarks inside. 
Up to now attempts to
model the nucleon as a soliton in the linear sigma model have
failed due to the instability of  the sigma solution arising from
the integration over the sea quarks. The evolution equation may be
helpful in finding such a soliton solution too. 

In our approximation we used lowest order in two cases to address the finite
density problem: the coupling constant $g$, was fixed during the
evolution and the wave function renormalization $Z$ was set to one. A
recent publication of the Wetterich group~\cite{Wett} uses a running
meson-quark coupling and obtains a solution with a mixed phase
extending from very low to very high density. In fact the evolution of
the coupling 
constant at finite density is not necessarily the same as at zero
density. The
equations for the wave function renormalization and 
the coupling are modified at finite density. Important effects such as
pion condensation~\cite{Yazaki} may appear after the wave function
renormalization. One therefore has to make more extended calculations to
know the full result for the linear sigma model at finite density.

In quark matter 
the linear sigma model with only attractive sigma mesons and pions
is not sufficient to  capture the nuclear physics.
Repulsive effects and confinement into nucleons play an
important role. To monitor the transition to the deconfined phase 
one needs an order parameter which keeps track of this transition.
Since the linear sigma model with quarks is already a hybrid model, 
it is not unnatural to include also nucleon degrees of freedom
explicitly and trace the transition of nucleon to quark degrees
of freedom directly.
This has been done in a separate paper~\cite{Schwenzer}.

\section*{Acknowledgments} 
One of the authors (G.P.) would like to thank J. Polonyi for
early discussions.
This work was supported in part by the US DOE grant DE-FG02-86ER40251
and the Hungarian grant OTKA F026622.
T.K. acknowledges DAAD for making his stay in Germany possible.
He is also grateful to H.-J. Pirner and C. Wetterich
 for hosting him during his stay in Heidelberg; he thanks all the members
 of Institut fuer Theoretische Physik for their warm hospitality to him.
This work is partly done during his sabbatical.  
\begin{appendix}

\section{Calculation of $\partial \Omega^{F}/\partial k$
 and $\partial \Omega^{B}/\partial k$}

In this appendix, we calculate the derivative 
 of the thermodynamical potentials in the heat kernel method.

\subsubsection*{The fermionic part of the flow equations:}
The fermion part of 
 the thermodynamical potential is written as 
\be
\label{app-b-1}
\Omega^F=N_cN_fI^F(k),
\ee
with
\be
I^F(k)=2T \sum_{n,\vec{q}}
\int\limits_0^\infty\! \f{d\tau}{\tau}\  
        f(k^2\tau)  
                e^{-\tau[(\nu_n+i\mu)^2+\vec{q}^2+g^2\phi^2]}.
\ee
Using the fact that
\be
\f{df(k^2\tau)}{dk}=-k^5\tau^3{\rm e}^{-k^2\tau},
\ee
and putting $E^2_{q,k}=g^2\phi^2+\vec{q}^2+k^2$,
the derivative  $dI^F/dk$ is evaluated to be
\be
\label{app-b-2}
\f{\partial I^F}{\partial k}&=& 
              -2k^5T \sum_{n,\vec{q}}\ \int\limits_0^\infty\! d\tau 
                \tau^2e^{-\tau[(\nu_n+i\mu)^2+E^2_{q,k}]}\nonumber 
                 \\
 \         &=& -2k^5T\biggl(\f{d}{dk^2}\biggl)^2\sum_{n,\vec{q}}
              \f{1}{(\nu_n+i\mu)^2+E^2_{q,k}}, \nonumber \\
           &=& -k^5T\biggl(\f{d}{dk^2}\biggl)^2\sum_{\vec{q}}
                 S^F(q,k)
\ee
with
\be
S^F(q,k)=\sum_{n}
              \biggl[\f{1}{(\nu_n+i\mu)^2+E^2_{q,k}}
                  +\f{1}{(\nu_n-i\mu)^2+E^2_{q,k}}\biggl],
\ee
where use of the fact has been made in the last equality 
that the sum w.r.t. the Matsubara
 frequencies $\nu_n$ does not change when the sign of the frequencies
 is changed.

Now the nice point is that the sum in $S(q,k)$ can be performed
 analytically, as follows;
\be
S^F(q,k)&=&\f{d}{dx^2}\sum_{n}
              \biggl[\ln[(\nu_n+i\mu)^2+x^2]
                  +\ln[(\nu_n-i\mu)^2+x^2]\biggl]_{x=E^2_{q,k}},
                      \nonumber \\ 
    &=& \f{d}{dx^2}\sum_{n}
              \biggl[\ln[\nu_n^2+(x-\mu)^2]
                  +\ln[\nu_n^2+(x+\mu)^2]\biggl]_{x=E^2_{q,k}}.
\ee
Then  evaluating the derivative, we have
\be
S^F(q,k)=\f{1}{E_{q,k}}\sum_{n=-\infty}^{\infty}
              \biggl[\f{E_{q,k}-\mu}{\nu_n^2+(E_{q,k}-\mu)^2}
                  +\f{E_{q,k}+\mu}{\nu_n^2+(E_{q,k}+\mu)^2}\biggl].
\ee
Now utilizing the formula
\be
\sum_{n=-\infty}^{\infty}\frac{1}{\nu_n^2+x^2}=\f{1}{2Tx}\tanh
 \f{x}{2T},
\ee
we end up with
\be
S^F(q,k)=\frac{1}{2TE_{q,k}}\left[\tanh\f{E_{q,k}-\mu}{2T}+
                \tanh\f{E_{q,k}+\mu}{2T}\right].
\ee
Inserting $S^F(q,k)$ into (\ref{app-b-2}),we have
\be
\f{\partial I^F}{\partial k}
 =-k^5\biggl(\f{d}{dk^2}\biggl)^2\int\f{d\vec{q}}{(2\pi)^3}
           \frac{1}{E_{q,k}}\biggl[1-n(E_{q,k})-\bar{n}(E_{q,k})\biggl],
\ee
where $n(x)$ ($\bar{n}(x)$) is the Fermi-Dirac distribution function, respectively
\be
n(x)=\frac{1}{{\rm e}^{(x-\mu)/T}+1}, \quad 
\bar{n}(x)=\frac{1}{{\rm e}^{(x+\mu)/T}+1}.
\ee

Now let us calculate the derivative w.r.t. $k^2$, which may be put
 into the integral; since $k^2$ appears only 
in the combination $q^2+k^2$,  the derivative can be converted to the
 one w.r.t. $q^2$. Then making a partial integration, we obtain
\be
\label{app-b-f}
\frac{\partial I^F}{\partial k^2}=\f{k^4}{8\pi^2}
       \frac{d}{dk^2} \int_0^{\infty}dq
    \frac{1}{E_{q,k}}(1-n(E_{q,k})-\bar{n}(E_{q,k})).
\ee
Inserting (\ref{app-b-f}) into (\ref{app-b-1}), we finally reach
  the formula presented in the text;
\be
\label{eq:fermi-1a}
\frac{\partial \Omega^F}{\partial k^2}=\frac{N_cN_f}{8\pi^2}k^4
       \frac{d}{dk^2} \int_0^{\infty}dq
    \frac{1}{E_{q,k}}(1-n(E_{q,k})-\bar{n}(E_{q,k})).
\ee

\subsubsection*{The bosonic part of the flow equations:}
Similarly, with the fermion part, the boson part  of 
 the thermodynamical potential involves the integral;
\be
\Omega^B(k)=-\f{T}{2} \sum_{n,\vec{q}}
\int\limits_0^\infty\! \f{d\tau}{\tau}\  
        f(k^2\tau)  
                e^{-\tau[\omega_n^2+\vec{q}^2+m^2]},
\ee
with $\omega_n$ being the Matsubara frequencies for bosons.
Here we shall confine ourselves to the  case 
where the boson has no chemical potential. The extention to 
the case  with finite chemical potential is easy.
 
The derivative w.r.t. $k$ can be performed as much the same way as the
 fermion part.
The only difference comes in with the formula
\be
\sum_{n=-\infty}^{\infty}\frac{1}{\omega_n^2+x^2}=\f{1}{2Tx}\coth
 \f{x}{2T}.
\ee
Thus we obtain,
\be
\f{\partial \Omega^B}{\partial k^2}
 =\f{k^4}{32\pi^2}\f{d}{dk^2}\int_0^{\infty} dq
           \frac{1}{E_{q,k}}\coth \f{E_{q,k}}{2T},
\ee
with $E^2_{q,k}=q^2+k^2+m^2$.

Performing the  derivative as before, we have
\be
\label{app-b-fb}
\frac{\partial \Omega^B}{\partial k^2} &=& \f{k^4}{64\pi^2} \left[
\frac{1}{m^2+k^2} \right. \\
&+& \left. 2\int_0^{\infty} dq \left\{ \f{1}{E^3_{q,k}}n_B(E_{q,k})
   +\f{1}{TE^2_{q,k}}n_B(E_{q,k})(1+n_B(E_{q,k})) \right\} \right],
        \nonumber
\ee
where $n_B(x)$ is the Bose-Einstein distribution function given by
\be
n_B(x)=\f{1}{{\rm e}^{x/T}-1}.
\ee
Taking the limit $T\rightarrow 0+$, we obtain
\be
\frac{\partial \Omega^B}{\partial k^2}=\f{k^4}{64\pi^2}
\frac{1}{m^2+k^2}.
\ee

\section{Connection between the mean field and the flow-eq. result}
\label{sec:connection}
Let us regard only the density dependent part of the evolution equations
(eq.~\ref{eq:flow-fermion}). If this part decouples from the meson
evolution, we
can integrate this equation analytically: 
\begin{eqnarray}
  \Omega^F_{k=0} & = & \int_{k_{initial}}^0 \frac{d\Omega_F}{dk^2}dk^2 \\
                 & = & \frac{N_c \mu}{8\pi^2} \int_{\mu^2-g^2\phi^2}^0
                       \frac{k^4 \; dk^2}{\left( k^2 + g^2\phi^2\right)
                                          \sqrt{\mu^2-k^2-g^2\phi^2}}.
\end{eqnarray}
The integral gives the following result:
\begin{eqnarray}
  \Omega^F_{k=0} &=&  - \frac{N_c}{8\pi^2} \left(
                   2\mu^3\sqrt{\mu^2-g^2\phi^2} - 4\mu g^2\phi^2
                   \sqrt{\mu^2-g^2\phi^2} -
                   \frac{2}{3} \mu \sqrt{\mu^2-g^2\phi^2}^3 \right. \nonumber\\
                   & &  \left. + 2 (g^2\phi^2)^2 \log{\frac{\mu +
                   \sqrt{\mu^2-g^2\phi^2}}{g\phi}}
                   \right). 
\end{eqnarray}
Since one can do the replacements,
\begin{eqnarray}
  \mu &=& \sqrt{k^2_F + g^2\phi^2} \nonumber \\
  k_F &=& \sqrt{\mu^2 - g^2\phi^2} \nonumber,
\end{eqnarray}
the potential takes the following form:
\begin{eqnarray}
  \Omega^F_{k=0} &=&   \frac{N_c}{4\pi^2} \left(2k_F \sqrt{k_F^2+g^2\phi^2}^3
                      - g^2\phi^2 \, k_F \sqrt{k_F^2+g^2\phi^2} \right. \nonumber \\
                 & & \left.    - g^4\phi^4  
                      \log{\frac{\sqrt{k_F^2+g^2\phi^2}+k_F}{g\phi}} \right)
                      - \frac{N_c}{4\pi^2} \frac{8}{3} \sqrt{k_F^2+g^2\phi^2}
                      k_F^3 \, .
\end{eqnarray}
Recalling the relation between $\Omega$ and $F$
\begin{equation}
  \Omega = F - 3\rho_B\mu \, ,
\end{equation}
we can identify the term in brackets with $F$, and this is indeed the fermionic
part of our mean field result eq.~(\ref{eq:energie}).

We can also do this independent check for the equation (\ref{eq:fflow}), but
because of the complicated form of our constraint equation this can not be done
analytically. Numerical integration of the density dependent part in
(\ref{eq:fflow}) reproduces again the mean field results.

\end{appendix}

\end{document}